\documentclass[aps,prd,twocolumn,showpacs,floats,floatfix,letterpaper,nofootinbib,superscriptaddress,]{revtex4-1}

\usepackage{amssymb,amsmath,latexsym,mathrsfs}
\usepackage{graphicx}
\usepackage{epsfig}

\newcommand{\mpl}{m_\mathrm{Pl}}
\newcommand{\srp}[1]{\lambda^{(#1)}_\mathrm{H}}
\newcommand{\srpl}{\srp{\ell}}

\begin{document}


\title{Impact of general reionization scenarios on extraction of 
inflationary parameters}

\author{Stefania Pandolfi}
\affiliation{Physics Department and ICRA, Universita' di Roma 
	``La Sapienza'', Ple.\ Aldo Moro 2, 00185, Rome, Italy}
\affiliation{Physics Department and INFN, Universita' di Roma 
	``La Sapienza'', Ple.\ Aldo Moro 2, 00185, Rome, Italy}
\author{Elena Giusarma}
\affiliation{IFIC, Universidad de Valencia-CSIC, 46071, Valencia, Spain}
\author{Edward W.\ Kolb}
\affiliation{Department of Astronomy \& Astrophysics, Enrico Fermi 
        Institute, and Kavli Institute for Cosmological Physics,
       	University of Chicago, Chicago, Illinois  60637, USA}
\author{Massimiliano Lattanzi$^1$}
\noaffiliation
\author{Alessandro Melchiorri$^2$}
\noaffiliation
\author{Olga Mena$^3$}
\noaffiliation
\author{Manuel Pe\~na$^3$}
\noaffiliation
\author{Asantha Cooray}
\affiliation{Center for Cosmology, Department of Physics \& Astronomy, 
	University of California, Irvine, California 92697, USA}
\author{Paolo Serra$^5$}
\noaffiliation

\begin{abstract}

Determination of whether the Harrison--Zel'dovich 
spectrum for primordial scalar perturbations is consistent with 
observations is sensitive to assumptions about the 
reionization scenario.  In light of this result, we revisit constraints 
on inflationary models using more general reionization scenarios. While 
the bounds on the tensor-to-scalar ratio are largely unmodified, when different 
reionization schemes are addressed, hybrid models are back into the 
inflationary game. In the general reionization picture, we reconstruct 
both the shape and amplitude of the inflaton potential. We find a broader 
spectrum of potential shapes when relaxing the simple reionization 
restriction. An upper limit of $10^{16}$ GeV to the amplitude of the 
potential is found, regardless of the assumptions on the reionization 
history.

\end{abstract}

\pacs{98.80.-k 95.85.Sz,  98.70.Vc, 98.80.Cq}

\maketitle

\section{Introduction}

The inflationary paradigm seems to be the ideal mechanism not only for
solving cosmological paradoxes such as the observed large-scale
smoothness and spatial flatness of our universe, but also for
providing the initial seeds for structure formation. The simplest
inflationary model makes use of a single scalar field $\phi$ (the
inflaton), which slowly evolves in a very shallow, nearly constant,
potential $V(\phi)$. The dynamics of slow roll gives rise to a
quasi-de Sitter phase of exponential expansion in the very early
universe. Since the slope of the spectrum is closely related to
derivatives of the field potential, slow-roll dynamics predicts that
the perturbation spectra should be very close to scale invariant,
although not \emph{exactly} so.  This implies that a quite general
inflationary prediction is that the power spectra of both scalar,
$P_\mathcal{R}$, and tensor, $P_T$, fluctuations can be well 
approximated by power laws, \emph{i.e.,}
\begin{equation}
P_\mathcal{R}(k)\propto k^{n-1}, \quad P_T(k)\propto k^{n_T},
\end{equation}
where the spectral indices $n$ and $n_T$ have very mild, if any,
dependence on the scale $k$. A scale-invariant scalar power spectrum
corresponding to the value $n=1$ is the model proposed by Harrison,
Zel'dovich, and Peebles \cite{hz}. In other words, from all the
considerations above, inflation predicts $n\simeq1$, but usually
$n\neq1$.\footnote{For a discussion of slow-roll inflation models with
$n=1$, see Ref.\ \cite{neq1}.}

A value of the spectral index $n$ slightly different from unity would 
strongly point to the inflationary paradigm as the mechanism responsible 
for providing the initial conditions for structure formation. In
addition, in many inflationary models the amplitude of gravitational
waves is proportional to $|n-1|$. Confirmation of a deviation from a
scale-invariant power spectrum would encourage the gravitational waves
hunters to keep searching for the detection of a nonzero tensor
amplitude.

The most recent analysis by the Wilkinson Microwave Anisotropy Probe 
(WMAP) team of their seven-year data \cite{wmap7} rule out the
Harrison-Zel'dovich (H--Z) primordial power spectrum at more than $3
\sigma$ when ignoring tensor modes: $n= 0.963\pm 0.012$.  
But this, as well as most other previously derived constraints 
from CMB data on cosmological parameters have assumed a ``sudden'' and 
complete reionization at a  single redshift $z_r$.  The reionization 
redshift, $z_r$, is taken to be in the range $4 < z_r < 32$, and the 
cosmological constraints are obtained after marginalization over $z_r$. 
The electron ionization fraction $x_{e}(z)$ is such that for 
$z\ll z_r$ $x_e(z) = 1$ ($x_e(z) = 1.08$ for $z < 3$ in order to take 
into account Helium recombination) and $x_e(z) = 2\times10^{-4}$ for 
$z > z_r$, \emph{i.e.,} joining the value after primordial recombination 
with a smooth interpolation.

The process of structure formation that led to gravitational collapse of 
objects in which the first stars formed are still subject to theoretical 
and observational uncertainties.  As these first sources began to illuminate 
their local neighborhoods, the HI present in the IGM was ``reionized.'' 
The end of the Dark Ages (the period between the end of CMB
recombination and the appearance of the first stars) remains to be 
explored and understood. 

There are two main effects on the CMB anisotropies produced by the
free electrons of the ionized gas: the first one washes out the
primary anisotropies of the temperature autocorrelation ($TT$)
spectrum. The damping of the $TT$ signal is quantified by the optical
depth parameter $\tau$, proportional to the column density of ionized
hydrogen.  Earlier reionization leads to a the larger suppression of
the $TT$ acoustic peaks. The second effect produces a damping and an
additional peak in the polarization autocorrelation spectrum ($EE$)
\cite{CMBPol}. The position of this new peak in the polarization
signal is proportional to the square root of the redshift at which the
reionization occurs, and its amplitude is proportional to the optical
depth. Since the precise details of reionization processes are
currently unknown, it is mandatory to explore the imprints of general
reionization histories on the CMB spectra.  In the standard, sudden
reionization scenario, the $EE$ spectrum depends exclusively on the
value of Thomson optical depth $\tau$. In turn, in extended
reionization schemes, the precise history of how the universe became
ionized affects the large-scale $EE$ power spectrum in a crucial way
\cite{mortonson}, and the power is transferred from larger to smaller
scales when considering that reionization processes could take place
in a non-negligible redshift (time) interval.

The major goal of this paper is to study how current constraints on the
scalar spectral index $n$ and the tensor-to-scalar ratio $r$ are
modified if the standard (``sudden'') reionization assumption is
relaxed.  In a precursor study we demonstrated that in a general
reionization scenario the Harrison-Zel'dovich spectrum ($n=1$) is
perfectly consistent with observations \cite{pandolfi}. In this study
we shall also include information from tensors modes, showing that
inflationary models that are ruled out in the sudden reionization
scheme are allowed in more general reionization scenarios.  We
also reconstruct both the shape and the amplitude of the inflationary
potential $V(\phi)$ allowed by current data in both sudden and general
reionization schemes.

The paper is organized as follows. Section \ref{sec:infl} summarizes the
results of inflationary theory relevant for our considerations.  A possible 
classification of different models of inflation is presented in Sec.\ 
\ref{sec:zoo}. The analysis method used here to derive the cosmological 
constraints is described in Sec.\ \ref{sec:ana}. Section \ref{sec:results} 
gives the resulting constraints on cosmological parameters and their 
implications for inflationary models. The inflationary potential 
reconstruction method and the results are presented in Sec.\ \ref{sec:MC}.  
We conclude in Sec.\ \ref{sec:concl}.

\section{Inflation and the Hamilton-Jacobi formalism} 
\label{sec:infl}

In this section we briefly review the dynamics of a scalar field in a
cosmological background. We assume a flat Friedmann-Robertson-Walker
(FRW) metric:
\begin{equation}
ds^2=dt^2 - a^2(t)[dr^2+r^2d\Omega^2]~,
\end{equation}
where $a(t)$ is the cosmological scale factor. In an FRW background, a
scalar field $\phi$ evolves under the action of potential $V(\phi)$
with equations of motion
\begin{eqnarray}
&& \ddot{\phi}+3H\dot{\phi}+ V'(\phi)=0 ~, \label{eq:phimot} \\
&& H^2=\frac{8\pi}{3\mpl^2}\left[\frac{1}{2}\dot\phi^2+V(\phi)\right]~, 
\label{eq:fried}
\end{eqnarray}
where $H\equiv \dot a/a$ is the Hubble parameter, dots and primes
denote derivatives with respect to cosmological time and to the scalar
field respectively, and $\mpl$ is the Planck mass. From the definition
of the Hubble parameter, it follows that
\begin{equation}
a(t) \propto e^{-N}= \exp\left[\int_{t_o}^t H(t)dt\right]~,
\end{equation}
where the number of $e$-folds $N$ is simply
\begin{equation}
N\equiv\int_t^{t_e} H(t)dt~,
\end{equation}
where $t_{e}$ refers to the end on inflation. The integration extrema are
chosen in such a way that $N=0$ coincides with the end of inflation. 

A very powerful way of describing the inflationary dynamics is given
by the Hamilton-Jacobi formulation of inflation. The basic idea is to
consider the scalar field $\phi$ itself to be the time variable; this
can be done as long as it varies monotonically with time. Then,
expressing the Hubble parameter as a function of the field,
$H=H(\phi)$, the equations of motion become
\begin{eqnarray}
&&\dot\phi=-\frac{\mpl^2}{4\pi} H'(\phi), \label{eq:HJ1}\\
&&[H'(\phi)]^2-\frac{12\pi}{\mpl^2}H^2(\phi)=-\frac{32\pi^2}{\mpl^4} V(\phi). 
\label{eq:HJ2}
\end{eqnarray}
The second of these equations is called the Hamilton-Jacobi
equation. Inflation takes place while the field is slowly rolling
towards a minimum of the potential, and the field energy density is
dominated by its potential energy.  More quantitatively, the slow-roll
approximation holds in the limit in which $\ddot{\phi}\ll
3H\dot{\phi}$ and $\dot{\phi}^2\ll V$, so that Eqs.\ (\ref{eq:phimot})
and (\ref{eq:fried}) become
\begin{eqnarray}
&&\dot{\phi}\simeq -\frac{V'(\phi)}{3H}~, \nonumber \\
&& H^2\simeq \frac{8\pi}{3\mpl^2}V(\phi)~. \label{eq:desit}
\end{eqnarray}
The validity of the slow-roll approximation is quite natural because 
the slope of the inflaton potential must be sufficiently shallow to 
drive inflation. For single-field inflation during the slow-roll phase, 
the kinetic energy of the field is negligible and the potential is nearly
constant:
\begin{equation}
\rho_{\phi}=V(\phi)+ \frac{\dot{\phi}^{2}}{2}\simeq V(\phi)\simeq 
\mathrm{const}.
\end{equation}
From Eq.\ (\ref{eq:desit}), we can see that this gives rise to a
(quasi-) de Sitter phase with $H$ almost constant. The amplitude of
the potential must be sufficiently large to dominate the energy
density of the Universe at that epoch.

The slow-roll approximation is consistent if both the slope and the
curvature of the potential are small (in units of the Planck mass)
when compared to the potential itself: $V', V'' \ll V$, or
equivalently if the so-called slow-roll parameters $\epsilon$ and
$\eta$ are much smaller than unity. The slow-roll parameters are
defined as 
\begin{equation} 
\epsilon \equiv \frac{\mpl^2}{4 \pi} \left [\frac{H'}{H}\right ]^2~,
\hspace{1cm} \eta \equiv \frac{\mpl^2}{4 \pi}\left[ \frac{H''}{H}\right]~. 
\label{eq:srdef}
\end{equation} 
When $V', V'' \ll V$, both $\epsilon$ and $\eta$ can be expressed
in terms of the potential and its derivatives as
\begin{eqnarray}
&&\epsilon \simeq \frac{\mpl^2}{16 \pi} \left ( \frac{V'}{V}\right )^2 \ll
 1\ ~,  \nonumber \\
&&\eta \simeq \frac{\mpl^2}{8 \pi}\left[ \frac{V''}{V}-\frac{1}{2}
\left(\frac{V'}{V}\right)^2\right] \ll1 ~.
\end{eqnarray}
The consistency of the slow-roll condition thus implies that 
$\epsilon,|\eta| \ll 1$. Using the definition of $\epsilon$, the 
Hamilton-Jacobi equation can be rewritten in the useful form
\begin{equation}
H^2(\phi)\left[1-\frac{1}{3}\epsilon({\phi})\right] = \frac{8\pi}{3\mpl^2}
 V(\phi)~.
\label{eq:hj}
\end{equation}

Inflation also provides a natural mechanism to generate the
inhomogeneities presently observed in the Universe. During inflation,
quantum fluctuations, inevitably present at small scales, are quickly
redshifted to scales much larger than the horizon size and then frozen
in as perturbations to the background metric. The perturbations
created during inflation can be of two types: scalar (or curvature)
perturbations, which couple to the matter stress-energy tensor, and
tensor perturbations (gravitational waves), which do not couple to
matter. The power spectrum of scalar perturbations (quantified as
perturbations in the Ricci scalar $\mathcal{R}$) is described by 
\begin{equation}
P_{\mathcal{R}}^{1/2}(k)=
\left(\frac{H^{2}}{2\pi|\dot\phi|}\right)_{k=aH}=
\left[\frac{H}{\sqrt{\pi}\mpl}\frac{1}{\sqrt{\epsilon}}\right]~,
\end{equation}
and its spectral index $n$ reads 
\begin{equation} 
n-1\equiv\frac{d\ln
P_{\mathcal{R}}}{d \ln k}~.  
\end{equation} 
The power spectrum of tensor fluctuation modes is given by 
\begin{equation} 
P_{T}^{1/2}(k)= \left
(\frac{4}{\sqrt{\pi}}\frac{H}{\mpl} \right )_{k=aH}~, 
\end{equation} 
again
evaluated when the mode $k$ crosses the horizon.
 
The ratio of the tensor-to-scalar perturbation is defined as 
\begin{equation}
\frac{P_{T}}{P_\mathcal{R}}\equiv r~, 
\end{equation} 
and, as in the scalar power spectrum case, one can write $P_{T}\propto
k^{n_{T}}$. The two spectral indices expressed in terms of the
slow-roll parameters are 
\begin{eqnarray} 
n & \simeq & 1-4\epsilon+2\eta~,\\ n_{T} & \simeq &-2\epsilon~, 
\end{eqnarray} 
and the tensor-to-scalar ratio $r$ is
\begin{equation} 
r\equiv 16 \epsilon~. 
\end{equation} 
The relations above are valid at first-order approximation in the
slow-roll parameters. Therefore, if primordial perturbations
originated from the dynamics of a slow-rolling scalar field, the
spectrum should not be exactly scale invariant. In fact, since the
slow-roll parameters $\epsilon$ and $\eta$ are small, but not
vanishing (in other words, since the potential is very close to flat
but not \emph{exactly} flat), we expect that $n\simeq 1$ but
nevertheless $n\ne 1$. A scale-invariant power spectrum corresponds to
the value $n=1$ is the aforementioned model proposed by
Harrison, Zel'dovich, and Peebles \cite{hz}. Given the fact that
$P_{\mathcal{R}}\propto k^{n-1}$, the spectral index can be thought as
a measure of the departure of the spectrum of the scalar perturbations
from an exactly scale-invariant power spectrum.

\section{ZOOLOGY OF INFLATIONARY MODELS}
\label{sec:zoo}

In this section we follow the classification of Kinney \emph{et al.}\
\cite{Kinney2006}. At lowest order in the slow-roll approximation
the relevant parameters to distinguish among inflationary models are
$n$ and $r$ \cite{dkk}. The different classes of models are
characterized by the relation between these two parameters, or
equivalently, by the relation between $\epsilon$ and $\eta$.  At
lowest order in the slow-roll approximation we can divide the
inflationary models into three general types: \emph{large-field},
\emph{small-field} and \emph{hybrid}. The boundary between large-field and
small-field models is represented by the so called \emph{linear} models.
\begin{itemize}
\item{Large-field models} 
are characterized by $-\epsilon<\eta\leq\epsilon$.  Popular examples of 
large-field models are $V(\phi)=\Lambda^{4}(\phi/\mu)^{p}$ and exponential 
potentials, $V(\phi)=\Lambda^{4}\exp(\phi/\mu)$.
\item{Small-field models}
are characterized by $\eta < -\epsilon$. They result from a generic potential 
of the form $V(\phi) = 
\Lambda^{4}[1-(\phi/\mu)^{p}]$, which can be understood as the lowest-order 
Taylor expansion of an arbitrary potential about the origin. 
\item{Hybrid models}
are characterized by $0<\epsilon<\eta$.   A generic hybrid potential is of 
the form $V(\phi)=\Lambda^{4} [1+(\phi/\mu)^{p}]$. 
\item{Linear models}
are on the boundary between large-field and 
small-field, and they are characterized for this reason by $\eta=-\epsilon$. 
The generic linear potential is of the form: $V(\phi)\propto\phi$.
\end{itemize}
With the above classification we can cover the entire $n$-$r$ plane and
derive constraints on the inflationary models directly from the
constraints on the $n$-$r$ plane that arise from cosmological
observations; see Sec.\ \ref{sec:results}.

\section{ANALYSIS METHOD}
\label{sec:ana}

We adopt two different methods for parametrization of the reionization
history. The first method, developed in Ref.\ \cite{mortonson}, is
based on principal components that provide a complete basis for
describing the effects of reionization on large-scale $E$-mode
polarization. Following Ref.\ \cite{mortonson}, one can parametrize
the reionization history as a free function of redshift by decomposing
$x_e(z)$ into its principal components:
\begin{equation}
x_e(z)=x_e^f(z)+\sum_{\mu}m_{\mu}S_{\mu}(z),
\end{equation}
where the principal components, $S_{\mu}(z)$, are the eigenfunctions
of the Fisher matrix that describes the dependence of the polarization
spectra on the electron ionization fraction $x_e(z)$, $m_{\mu}$ are
the amplitudes of the principal components for a particular
reionization history, and $x_e^f(z)$ is the WMAP fiducial model at
which the Fisher matrix is computed and from which the principal
components are obtained. In what follows we use the publicly available
$S_{\mu}(z)$ functions and vary the amplitudes $m_{\mu}$ for
$\mu=1,...,5$ for the first five eigenfunctions. Hereafter we refer to
this method as the MH (Mortonson-Hu) case.

In a second approach to a general reionization prescription we
employ a different parametrization,
sampling the evolution of the ionization fraction $x_e$ as a function
of redshift $z$ at seven points ($z = 9, 12, 15, 18, 21, 24, \textrm{and}\ 
27$), and interpolating the value of $x_{e}(z)$ between them with a cubic
spline. For $30 < z$ we fix $x_e = 2\times10^{-4}$ as the value of
$x_e$ expected before reionization (and after primordial
recombination), while $x_e = 1$ for $3 < z < 6$ and $x_e = 1.08$ for
$z < 3$ in order to be in agreement with both Helium ionization and
Gunn-Peterson test observations. This approach is very similar to the
one used in Ref.\ \cite{Lewis}, and we will refer to it as the LWB
(Lewis-Weller-Battye) case.

We then modified the Boltzmann CAMB code \cite{camb},
incorporating the two generalized reionization scenarios and extracted
cosmological parameters from current data using a Monte Carlo Markov
Chain (MCMC) analysis based on the publicly available MCMC package
\texttt{cosmomc} \cite{Lewis:2002ah}.

We consider here a flat $\Lambda$CDM universe described by a set of
cosmological parameters
\begin{equation}
 \label{parameter}
 \{\omega_b,\omega_c, \Theta_s, n, \log[10^{10}A_{s}], r, n_{run} \},
\end{equation}
where $\omega_b\equiv\Omega_bh^{2}$ and $\omega_c\equiv\Omega_ch^{2}$
are the physical baryon and cold dark matter densities relative to the
critical density, $\Theta_{s}$ is the ratio between the sound horizon
and the angular diameter distance at decoupling, $A_{s}$ is the
amplitude of the primordial spectrum, $n$ is the scalar spectral
index, $r$ is the tensor-to-scalar ratio, and $n_{run}\equiv d n/d \ln
k$ is the running of the scalar spectral index:
\begin{equation}
 \Delta^2_{\cal R}(k) = \Delta^2_{\cal R}(k_0)
\left(\frac{k}{k_0}\right)^{n(k_0)-1+\frac12\ln(k/k_0)dn/d\ln k}~.
\end{equation} 
Here, $k_0 = 0.002 \textrm{ Mpc}^{-1}$ is the pivot scale.

The extra parameters needed to describe reionization are the five
amplitudes of the eigenfunctions for the MH case, or the seven
amplitudes in the seven bins for the LWB case, and one single common
parameter, the optical depth, $\tau$, for the sudden reionization
case.

Our basic data set is the seven--year WMAP data \cite{wmap7}
(temperature, polarization, and tensor modes) with the routine for
computing the likelihood supplied by the WMAP team. Together with the
WMAP data, we also augment the WMAP7 data with the CMB data sets from
BOOMERanG \cite{boom03}, QUAD \cite{quad}, ACBAR \cite{acbar}, and
BICEP \cite{bicep}. For all these experiments we marginalize over a
possible contamination from the Sunyaev-Zel'dovich component,
rescaling the WMAP template at the corresponding experimental
frequencies. We therefore consider two cases: we first analyze the
WMAP data alone, referring to it as to the ``WMAP7'' case, and we then
include the remaining CMB experiments (``CMB-ALL'').

\section{RESULTS}
\label{sec:results}

Table \ref{table1} summarizes the main results of the analysis for
different cosmological data sets, showing the constraints on $n$ and
$r$ for the MH, LWB, and sudden reionization schemes. When the sudden
reionization assumption is relaxed, the mean values of $n$ and $r$ tend to 
shift to higher values. The shift in $n$ was already noted in the 
previous paper \cite{pandolfi}. The importance of this shift is that in a 
general reionization scheme the H--Z spectrum is perfectly consistent.  
Notice, however, that the presence of tensors and/or a running spectral index 
in the analysis allows for a H--Z spectrum even in the sudden reionization 
scheme (at a confidence level (c.l.) corresponding to $2\sigma$).  
Nevertheless,  in the case of the MH reionization scenario, without running 
of the index, the best fit for the scalar spectral index is already higher 
than one at 68\% c.l.  In a general reionization scenario the allowed values 
of $r$ also shift to higher values.  When additional data from other CMB 
probes are added to the WMAP7 data, the constraints on $n$ and $r$ are 
shifted back toward lower values.  In summary, in the MH reionization 
case ignoring running of the spectral index, using WMAP7 data the H--Z
spectrum ($n=1$) is very close to the best fit value, and inside
the 68\% c.l.\ for the case CMB-ALL.

\begin{table*}
\begin{center}
\begin{ruledtabular}
\begin{tabular}{l|ccc|ccc|c}
& \multicolumn{3}{c|}{WMAP7} & \multicolumn{3}{c|}{CMB-ALL} & Planck \\ \hline
& Sudden & MH & LWB & Sudden & MH & LWB & Sudden \\ \hline \hline
& & & & & & &\\
\multicolumn{8}{c}{no running of scalar spectral index} \\[0.2cm]
$n\ (68 \%\ \textrm{c.l.)}$			&
$0.987\pm0.020$					&
$1.001\pm0.027$					&
$0.992\pm0.021$					&
$0.974\pm0.016$					&
$0.985\pm0.020$					&
$0.977\pm0.017$					&
$0.960\pm0.004$					\\[0.2cm]
$n\ (95 \%\ \textrm{c.l.)}$			&
$n\le 1.031$					&
$n\le 1.067$					&
$n\le 1.039$					&
$n\le 1.007$					&
$n\le 1.026$					&
$n\le 1.012$					&
$n\le 0.968$					\\[0.2cm]
$r\ (68 \%\ \textrm{c.l.)}$			&
$0.142\pm0.116$					&
$0.141\pm0.119$					&
$0.149\pm0.115$					&
$0.095\pm0.079$					&
$0.101\pm0.085$					&
$0.103\pm0.087$					&
$0.053\pm0.022$					\\[0.2cm]
$r\ (95 \%\ \textrm{c.l.)}$			&
$r\le 0.373$					&
$r\le 0.376$					&
$r\le 0.371$					&
$r\le 0.251$					&
$r\le 0.266$					&
$r\le 0.275$					&
$r\le 0.093$					\\[0.2cm]
\multicolumn{8}{c}{running of scalar spectral index} \\[0.2cm]
$n\ (68 \%\ \textrm{c.l.)}$			&
$1.067\pm0.062$					&
$1.080\pm0.065$					&
---						&
$1.094\pm0.052$					&
$1.106\pm0.054$					&
---						&
---						\\[0.2cm]
$n\ (95 \%\ \textrm{c.l.)}$			&
$n\le 1.192$					&
$n\le 1.207$					&
---						&
$n\le 1.197$					&
$n\le 1.222$					&
---						&
---						\\[0.2cm]
$r\ (68 \%\ \textrm{c.l.)}$			&
$0.191\pm0.154$					&
$0.200\pm0.158$					&
---						&
$0.181\pm0.141$					&
$0.185\pm0.140$					&
---						&
---						\\[0.2cm]
$r\ (95 \%\ \textrm{c.l.)}$			&
$r\le 0.497$					&
$r\le 0.515$					&
---						&
$r\le 0.451$					&
$r\le 0.445$					&
---						&
---						\\[0.2cm]
$n_\textrm{run}\ (68 \%\ \textrm{c.l.})$ 	&
$-0.040\pm0.029$				&
$-0.036\pm0.031$				&
---						&
$-0.056\pm0.021$				&
$-0.058\pm0.022$				&
---						&
---						\\[0.2cm]
& & & & & & &\\
\end{tabular}
\end{ruledtabular}

\caption{Constraints for different data sets on $n$, $r$, and
$r_\textrm{run}$ in different reionization scenarios with and without
the running of the scalar spectral index $n$.}

\label{table1}
\end{center}
\end{table*}

\begin{figure*}
 \centering
\includegraphics[width=8.5cm]{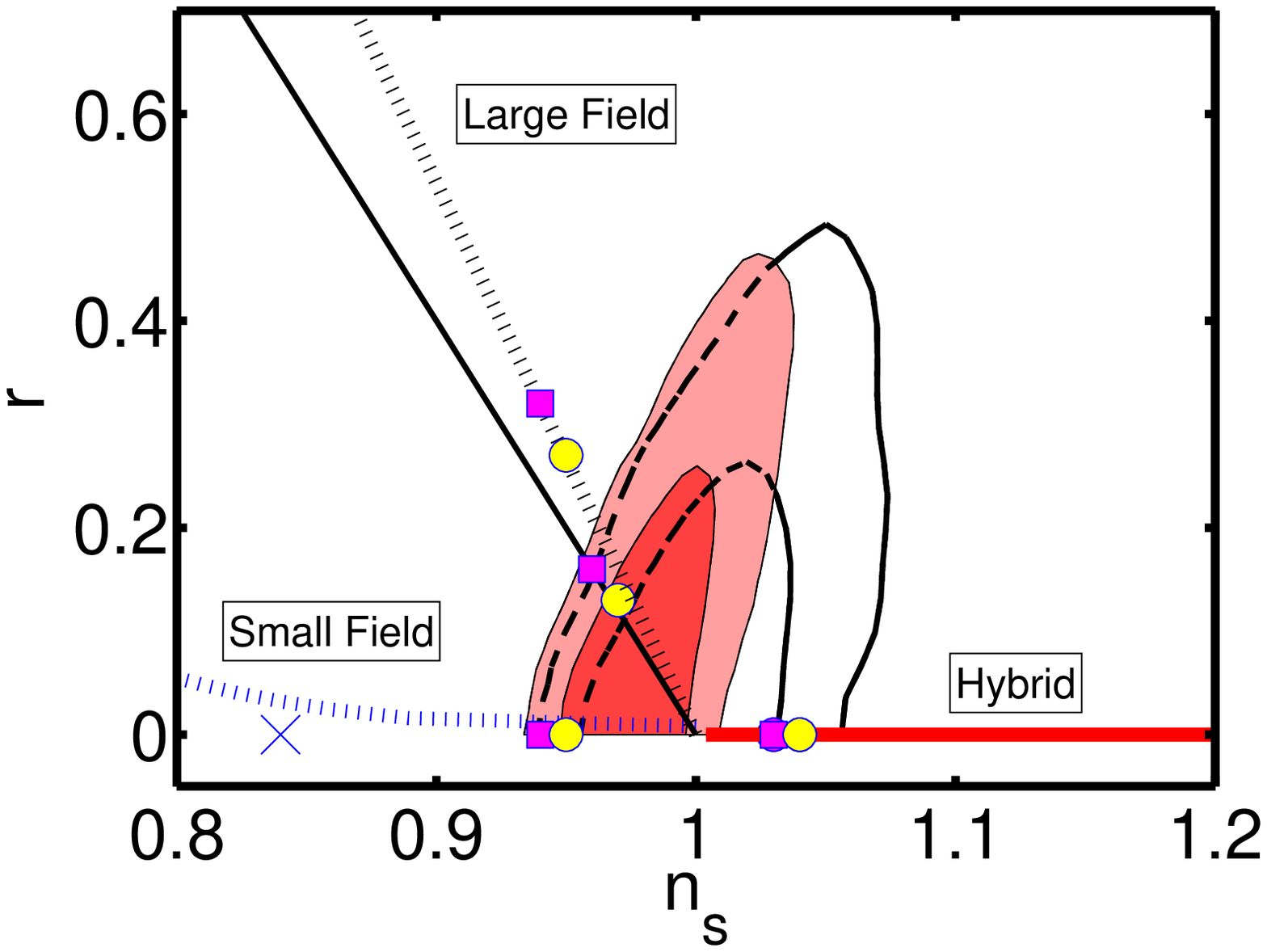}
\includegraphics[width=8.5cm]{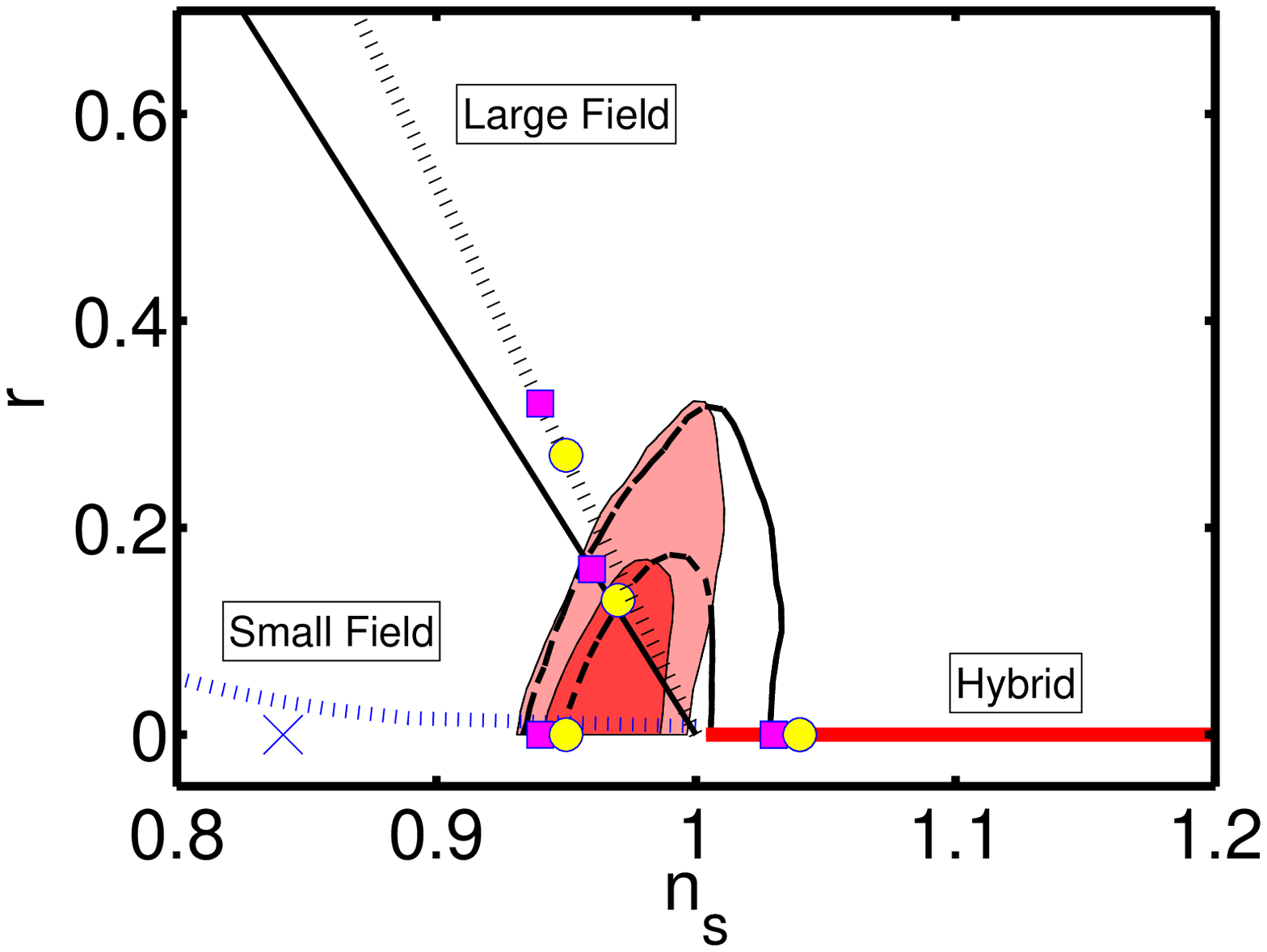}
\caption{Two-dimensional contour plots at the 68\% and 95\% confidence
levels without running of the scalar spectral index for the WMAP7 data
(left figure) and the CMB-ALL data set (right figure). Shaded contours
correspond to the sudden reionization approximation, while open
contours model reionization as MH.  The dark solid (dashed) lines
refer to large-field models with $p=2$ ($p=4$). The lighter cross
(dashed) curves depict small-field models with $p=2$ ($p=4$). The
solid horizontal line that basically coincides with the $x$ axis
depicts hybrid models with $p=2$ (the $p=4$ case basically overlaps
the $p=2$ case).  The filled circles (squares) denote the points in
the parameter space for which the number of $e$-folds $N$ is equal to
60 (50).}
\label{fig:inflation_potential_norun_lab}
\end{figure*}
\begin{figure*}
 \centering
\includegraphics[width=8.3cm]{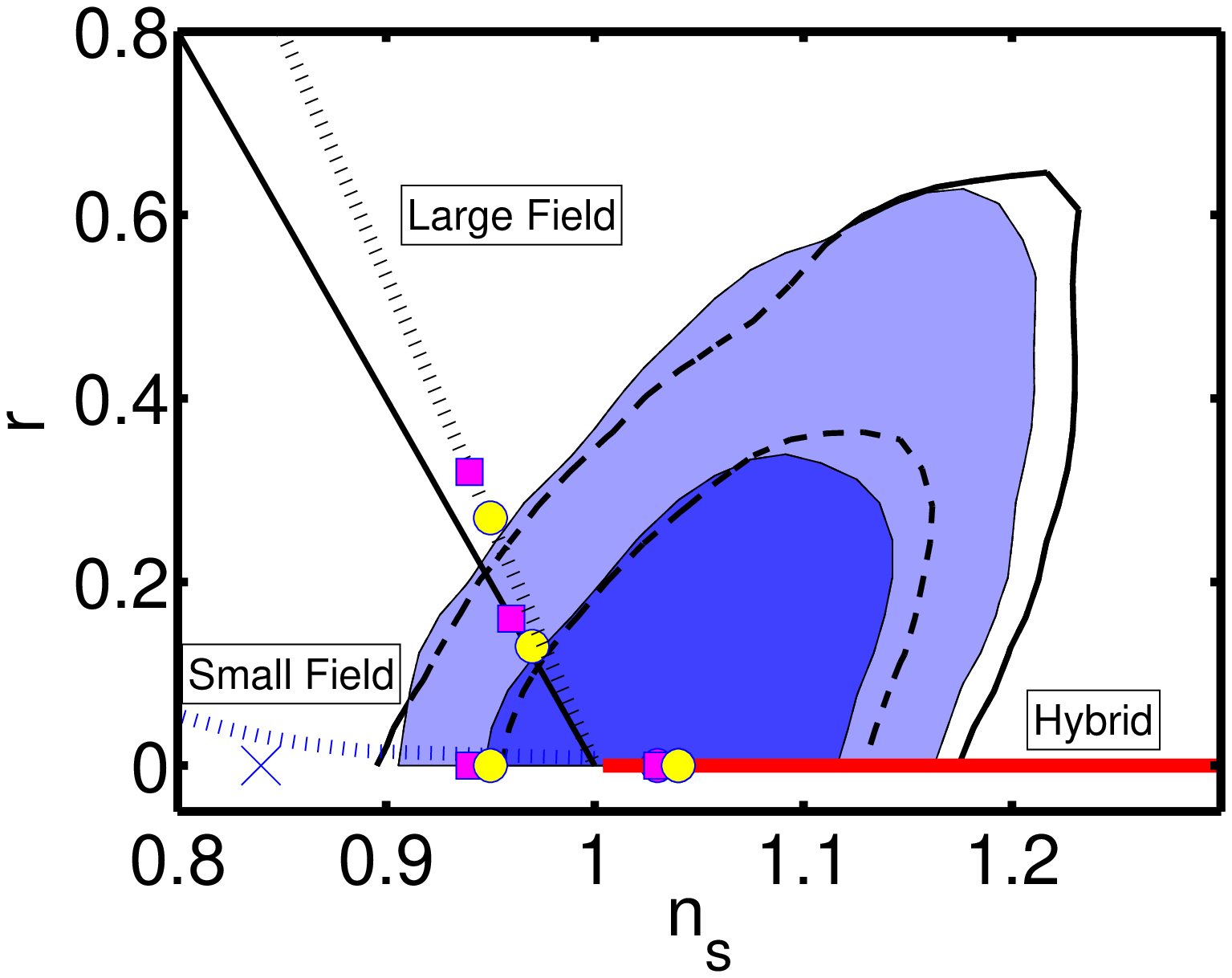}
\includegraphics[width=8.3cm]{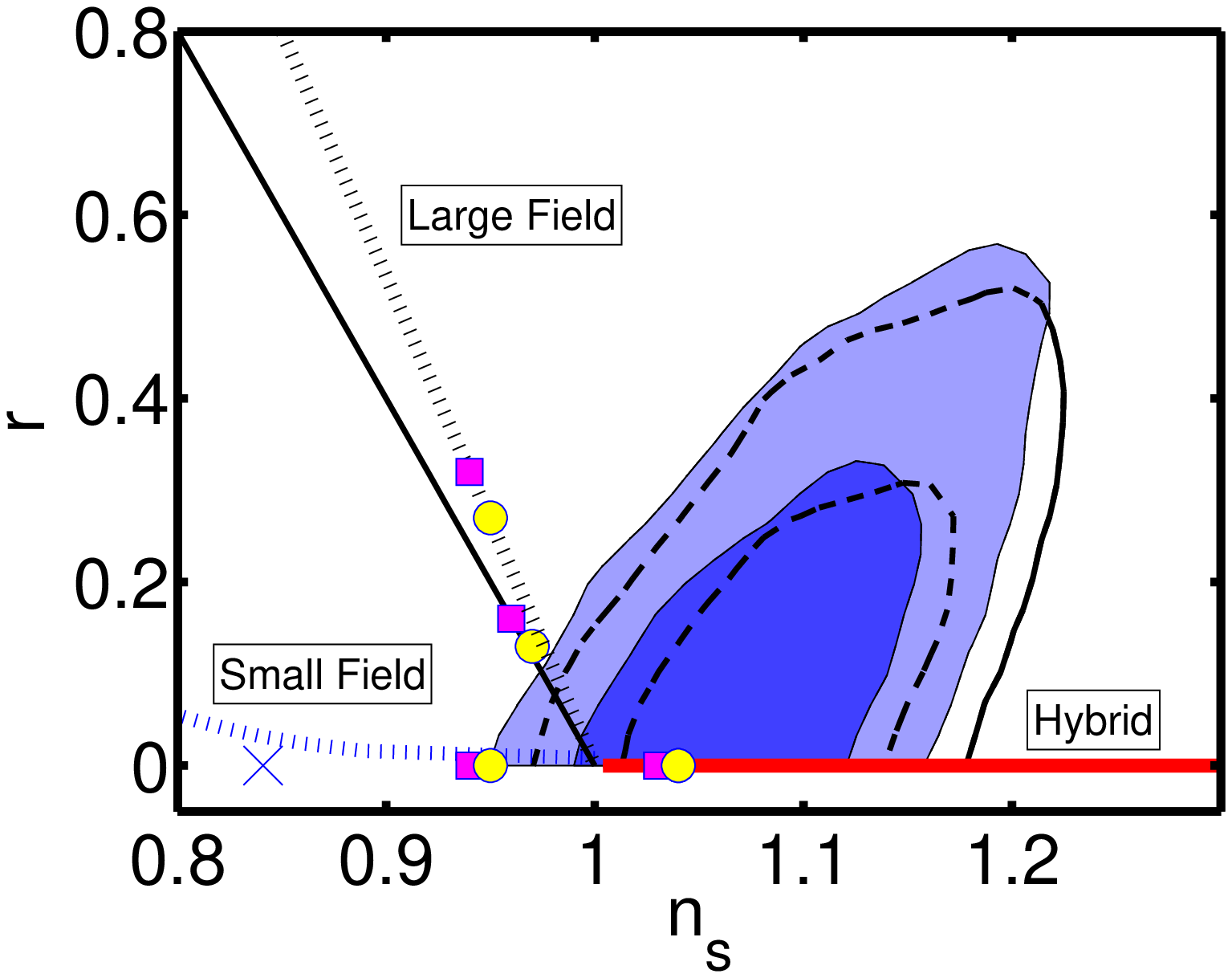}
\caption{Two-dimensional contour plots at the 68\% and 95\% confidence
levels with running of the scalar spectral index for the WMAP7 data
(left figure) and the CMB-ALL data set (right figure).  The key for
the figures is the same as in Fig.\ \ref{fig:inflation_potential_norun_lab}.}
\label{fig:inflation_potential_run_lab}
\end{figure*}

The values for the tensor-to-scalar ratio $r$ and the running of the
spectral index $n_{run}$ at 95\% c.l.\ are slightly higher considering
a general reionization scenario. However, their 68\% c.l.\ constraints
barely change when the reionization history is modified, as expected,
due to the large uncertainties on $r$ and $n_{run}$.

The shift induced on allowed values of inflationary parameters $n$ and 
$r$ by different assumptions for the reionization history is important 
for the subsequent constraints on inflationary models. To study this, 
we have reconstructed the relation between $n$ and $r$ in the different 
classes of models described in the previous section, and we have plotted 
these relations in the $n$-$r$ plane, together with the cosmological 
constraints.

Figure \ref{fig:inflation_potential_norun_lab} depicts the 68\% and
95\% c.l.\ allowed contours by the WMAP7 data and the CMB-ALL data sets
without running of the scalar spectral index for different assumptions
of the reionization history.  The indicated contours denote the
allowed regions when tensor modes are included in the analysis, and when 
the reionization is assumed to be sudden and when using the MH procedure 
(see the figure caption for details).

Figure \ref{fig:inflation_potential_run_lab} is the same as Fig.\ 
\ref{fig:inflation_potential_norun_lab} but now allowing for a running of 
the scalar spectral index.

Following Ref.\ \cite{Kinney2006} we can easily develop the different
expressions concerning the $n$-$r$ parameter space. For instance, for
large-field models, with a polynomial potential $V \propto \phi^p$, the
relation among these parameters is
\begin{equation}
n=1-\frac{r}{8}\left ( 1+\frac{2}{p} \right)~.
\end{equation}
The dark lines in Figs.\ \ref{fig:inflation_potential_norun_lab} and
\ref{fig:inflation_potential_run_lab} refer to this relation for
quadratic ($p=2$) and quartic ($p=4$) potentials. It is
straightforward to relate $r$ with $N$ (the number of $e$-foldings before
the end of inflation):
\begin{equation}
r=4p/N~,
\end{equation}
which allows us to draw points with $N=50$ (squares) and $N=60$ (circles) in 
Figs.\ \ref{fig:inflation_potential_norun_lab} and
\ref{fig:inflation_potential_run_lab}.

Similarly, we can relate $n$ to $r$, and both of them in terms
of $N$ for both small-field and hybrid models. For small-field models,
the generic potential we are using is of the form $V(\phi) = \Lambda^4
[ 1-(\phi/\mu)^p]$.  Typically in these models the slow-roll parameter
$\epsilon$ (and hence, $r$) is close to zero. The spectral index can
be written as
\begin{equation}
n\simeq1-\frac{p(p-1)}{4 \pi} \frac{m_{Pl}^2}{\mu^p} \left[ 
\frac{\pi\mu^{2p}}{m_{Pl}^2p^2} \ r \right]^{(p-2)/(2p-2)} ~.
\end{equation}

It is straightforward to see that for $p=2$
\begin{equation}
 n\simeq1-\left (\frac{1}{2\pi} \right )\left (\frac{m_{Pl}}{\mu}\right)^2~,
 \end{equation}
while for $p=4$ we have
 \begin{equation}
 n\simeq1-\frac{3}{\pi}\left (\frac{\pi m_{Pl}^4r}{16\mu^4}\right)^{1/3}
\approx  1-\frac{3}{N}~.
\end{equation}
Figs.\
\ref{fig:inflation_potential_norun_lab} and
\ref{fig:inflation_potential_run_lab} also contain the small-field model 
case, depicted by cross for $p=2$ and by the indicated dashed curve for 
$p=4$ (assuming $\mu\approx m_{Pl}$).

\begin{figure}
\centering
\includegraphics[width=8.0cm]{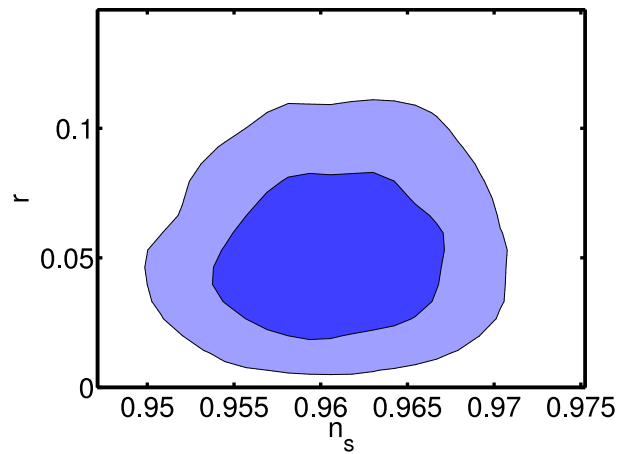}
\caption{The 68\% and 95\% c.l.\ constraints forecast on the $n$ vs.\ $r$ 
plane from  Planck mock data for $n=0.96$ and $r=0.05$ and sudden reionization 
(dark  contour), and  (wrongly) fitted assuming MH reionization (light 
contour).} \label{fig:planck_r0.05}
\end{figure}

For hybrid models, the potential chosen is
$V(\phi)=\Lambda^4[1-\alpha(m_{Pl}/\phi)^p]$, based on potentials
generated in dynamical SUSY breaking models \cite{lyth}. As in small-field 
models, the tensor-to-scalar ratio $r$ is negligible. The
expression for $n$ is given by
\begin{equation}
n\approx 1+2 \frac{(p+1)}{(p+2)(N_\mathrm{tot}-N)}~,\label{n_hybrid}
\end{equation}
where $N_\mathrm{tot}$ is the \emph{total} number of $e$-foldings
(chosen to be $100$ in this example). Notice that Eq.\
(\ref{n_hybrid}) indicates that the power spectrum in these sort
of models is blue ($n>1$). Indeed, in the sudden reionization scenario
with negligible running of the spectral index these hybrid models are
highly disfavored; in more general reionization schemes such models
are allowed by WMAP7 data; see Fig.\
\ref{fig:inflation_potential_norun_lab}. When more CMB data sets are
included in the analysis, hybrid inflation models with a blue tilt are
again disfavored at 95\% c.l., even in the more general reionization
scenarios considered here; see the CMB-ALL part of Fig.\
\ref{fig:inflation_potential_norun_lab}.  When a running scalar spectral 
index is allowed, hybrid models are perfectly compatible with data,
regardless of the assumptions about the reionization processes; see Fig.\
\ref{fig:inflation_potential_run_lab}.

The LWB reionization scheme leads to very similar constraints to those
of MH parametrization on the $n-r$ plane (albeit slightly closer to
the sudden reionization case). Indeed, $n$ is constrained to be red at the
68\% c.l.\ in the CMB-ALL case, but the H--Z model is still consistent
with data within two standard deviations.

We also forecast future constraints from the Planck experiment with
the specifications of Ref.\ \cite{bluebook}, assuming that $n=0.96$ and
$r=0.05$ and sudden reionization. If the data is (wrongly) fitted
assuming a more general reionization scenario (MH reionization, for
instance), the constraints that one would obtain on the $n-r$ plane
are shown in Fig.\ \ref{fig:planck_r0.05}. Notice that Planck will be
able to tell $n\neq 1$ at a very high confidence level even if the
precise details of the reionization processes are unknown. Planck data
will also be sensitive to the tensor-to-scalar ratio at the $95\%$
c.l.\ for $r \geq 0.05$.

\section{Monte Carlo Reconstruction of the inflationary potential}
\label{sec:MC}

In this section we describe the technique known as Monte Carlo
reconstruction, a stochastic method for inverting observational
constraints to obtain an ensemble of inflationary potentials
compatible with observations. The method is described in more detail
in Refs.\  \cite{Kinney:2002qn,Easther:2002rw,Kinney:2003uw}.

The slow roll parameters $\epsilon$ and $\eta$ already have been 
defined in Sec.\ \ref{sec:infl}; see Eq.\ (\ref{eq:srdef}). These two
parameters are related to the observables $n$ and $r$ by the formulae
given in Sec.\ \ref{sec:infl}, valid to first order in the slow-roll
approximation. We find it convenient to use the parameter $\sigma\equiv
2\eta-4\epsilon$ in place of $\eta$; the advantage is that to first
order in slow roll, $\sigma \simeq n-1$.

Taking higher derivatives of $H$, one can construct an infinite hierarchy of
slow-roll parameters \cite{Liddle:1994dx}:
\begin{equation}
\srpl\equiv \left(\frac{\mpl^2}{4\pi} \right)^\ell 
\frac{(H')^{\ell-1}}{H^\ell}\frac{d^{(\ell+1)}H}{d\phi^{(\ell+1)}}
\end{equation}

The evolution of the slow-roll parameters is described by the following set 
of equations  \cite{Hoffman:2000ue,Schwarz:2001vv,Kinney:2002qn}:
\begin{eqnarray}
\frac{d\epsilon}{dN} & = &\epsilon(\sigma+2\epsilon), \label{eq:epsev} \\
\frac{d\sigma}{dN} & = &-5\epsilon\sigma -12\epsilon^2+2\srp{2}, 
\label{eq:sigmaev} \\
\frac{d\srpl}{dN} & =& \left[ (\ell-1)\frac{\sigma}{2}+(\ell-2) \epsilon 
\right] \srpl+\srp{\ell+1}, \label{eq:srpev}
\end{eqnarray}

Given a solution to these equations, the observable quantities,
\emph{i.e.,} the scalar spectral index $n$, its running $n_{run}$, and the
tensor-to-scalar ratio $r$, can be evaluated. To second order in slow
roll, these are given by \cite{Stewart:1993bc,Liddle:1994dx}:
\begin{eqnarray}
r & = & 16\epsilon[1-C(\sigma+2\epsilon)], \nonumber \\
n-1 & = & \sigma-(5-3C)\epsilon^2-\frac{1}{4}(3-5C)\sigma\epsilon 
\nonumber \\
& & + \frac{1}{2}(3-C)\srp{2}, \nonumber \\
n_{run} &
= &-\left(\frac{1}{1-\epsilon}\right) \frac{dn}{dN}~,
\end{eqnarray}
where $C\equiv 4(\ln2+\gamma)-5=0.08145$ and $\gamma\simeq 0.577$ is 
Euler's constant.

The solution to Eqs.\ (\ref{eq:epsev}-\ref{eq:srpev}) also allows one to 
reconstruct the form of the potential $V(\phi)$ 
\cite{Hodges:1990bf,Copeland:1993ie,AyonBeato:2000xx,Easther:2002rw}. 
In fact, from the Hamilton--Jacobi equation (see Eq.\ (\ref{eq:srdef}))
\begin{equation}
V(\phi)=\left(\frac{3 \mpl^2}{8\pi}\right) H^2(\phi)
\left[1-\frac{1}{3}\epsilon(\phi)\right] . 
\end{equation}
Once $\epsilon(N)$ is known from the solution of Eqs.\
(\ref{eq:epsev}-\ref{eq:srpev}), $H(N)$ can be determined from
\begin{equation}
\frac{1}{H}\frac{dH}{dN} = \epsilon.
\end{equation}
The solution to the above equation allows one then to obtain $V(N$) up to a 
normalization constant; this is fixed by the normalization of the Hubble 
parameter that enters the above equation as a integration constant. We will 
return on this later.

Finally, in order to obtain $\phi(N)$, we note that Eq.\ (\ref{eq:HJ1}) and
$dN/dt=-H$ together imply that
\begin{equation}
\frac{d\phi}{dN}=\frac{\mpl^2}{4\pi}\frac{H'}{H}=\frac{\mpl}{2\sqrt\pi}
\sqrt{\epsilon} ,
\end{equation}
where it should be implicitly understood that there is a sign ambiguity in 
the last equality since it should have the same sign as
$H'(\phi)$. Since we do not know in advance the sign of $H'$, we
should consider the two cases separately.  However, it is easy to see
that they are related by the transformation $\phi\rightarrow -\phi$.
For this reason, in the following we just consider the case
$d\phi/dN<0$ (i.e, the value of $\phi$ grows as inflation goes on). Once $\phi(N)$ is obtained from the solution to the
above equation, it can be inverted to obtain $N(\phi)$ and finally
$V(\phi$).  However, the value of $\phi(N)$ can be known only up to an
additive integration constant. We fix the latter so that $\phi=0$ at
the beginning of inflation.

\begin{table*}
\begin{center}
\begin{ruledtabular}
\begin{tabular}{l|cc|cc}
& \multicolumn{2}{c|}{WMAP7} & \multicolumn{2}{c}{CMB-ALL} \\ \hline
& Sudden & MH & 
Sudden & MH \\ \hline \hline 
& & & & \\
$n\ (99 \%\ \textrm{c.l.})$			& 
$0.913 \le n \le 1.221$  			& 
$0.918 \le n \le 1.234$ 			& 
$0.968 \le n \le  1.216$			&  
$0.984 \le n \le 1.260$				\\[0.2cm]
$r\ (99 \%\ \textrm{c.l.})$ 			& 
$r\le0.665$ 					&  
$r\le 0.688$					&  
$r\le 0.551$					&  
$r\le 0.619$ 					\\[0.2cm]
$n_\textrm{run}\ (99 \%\ \textrm{c.l.})$ 	& 
$-0.098 \le n_{run} \le +0.037 $ 		&  
$-0.098 \le n_{run} \le + 0.050$		&  
$ -0.099 \le n_{run} \le-0.001$			&  
$-0.117 \le n_{run} \le -0.001$			\\
& & & & \\
\end{tabular}
\end{ruledtabular}
\caption{Observational limits (at 99\% confidence level) used to 
constrain the inflationary potential for the four cases considered in
reconstruction.}
\label{table2}
\end{center}
\end{table*}

In order to calculate an ensemble of potentials that are compatible with
observations, we proceed in the following way:
\begin{enumerate}
\item Choose random initial values for the inflationary parameters in the
following ranges:
\begin{eqnarray*}
N &=&[40,\,70] \\
\epsilon&=&[0,\,0.8]  \\
\sigma&=&[-0.5,\,0.5] \\
\srp{2}&=&[-0.05,\,0.05] \\
\srp{3}&=&[-0.005,\,0.005] \\
& \vdots & \\
\srp{6}&=&0.
\end{eqnarray*}
truncated at $M=6$.
\item Evolve forward in time ($dN<0$) until either \emph{(a)} inflation ends 
($\epsilon>1$), or \emph{(b)} the evolution reaches a late-time fixed point
($\epsilon=\srpl=0$, $\sigma=\mathrm{const}$).
\item In case \emph{(a)}, evolve $N$ $e$-folds backwards in time from the end
of inflation and calculate the observables $n-1$, $r$, and the running
$n_{run}$ at that time; in case \emph{(b)}, calculate the observables at the
time the evolution reaches the fixed point.
\item Repeat the above procedure $N_{MC}$ times.
\item Choose a window of acceptable values for the observables $n-1$, $r$, and 
the running $n_{run}$, and then extract from the $N_{MC}$ models those 
that satisfy the observational constraints.
\item Reconstruct the potential for these models, following the procedure 
described above.
\end{enumerate}

\begin{figure*}
\begin{center}
\includegraphics[width=0.45\linewidth,keepaspectratio,clip]
{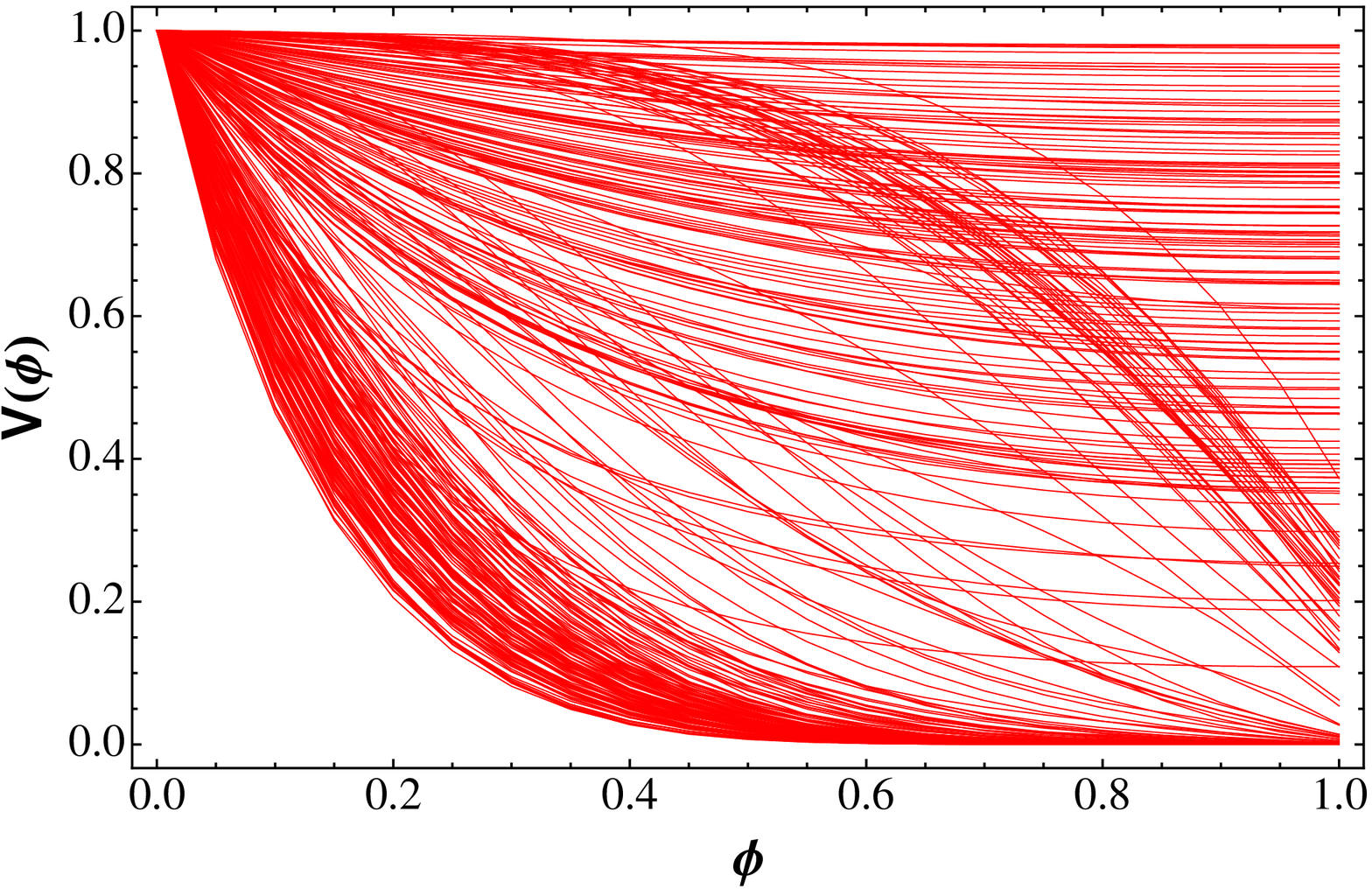}
\includegraphics[width=0.45\linewidth,keepaspectratio,clip]
{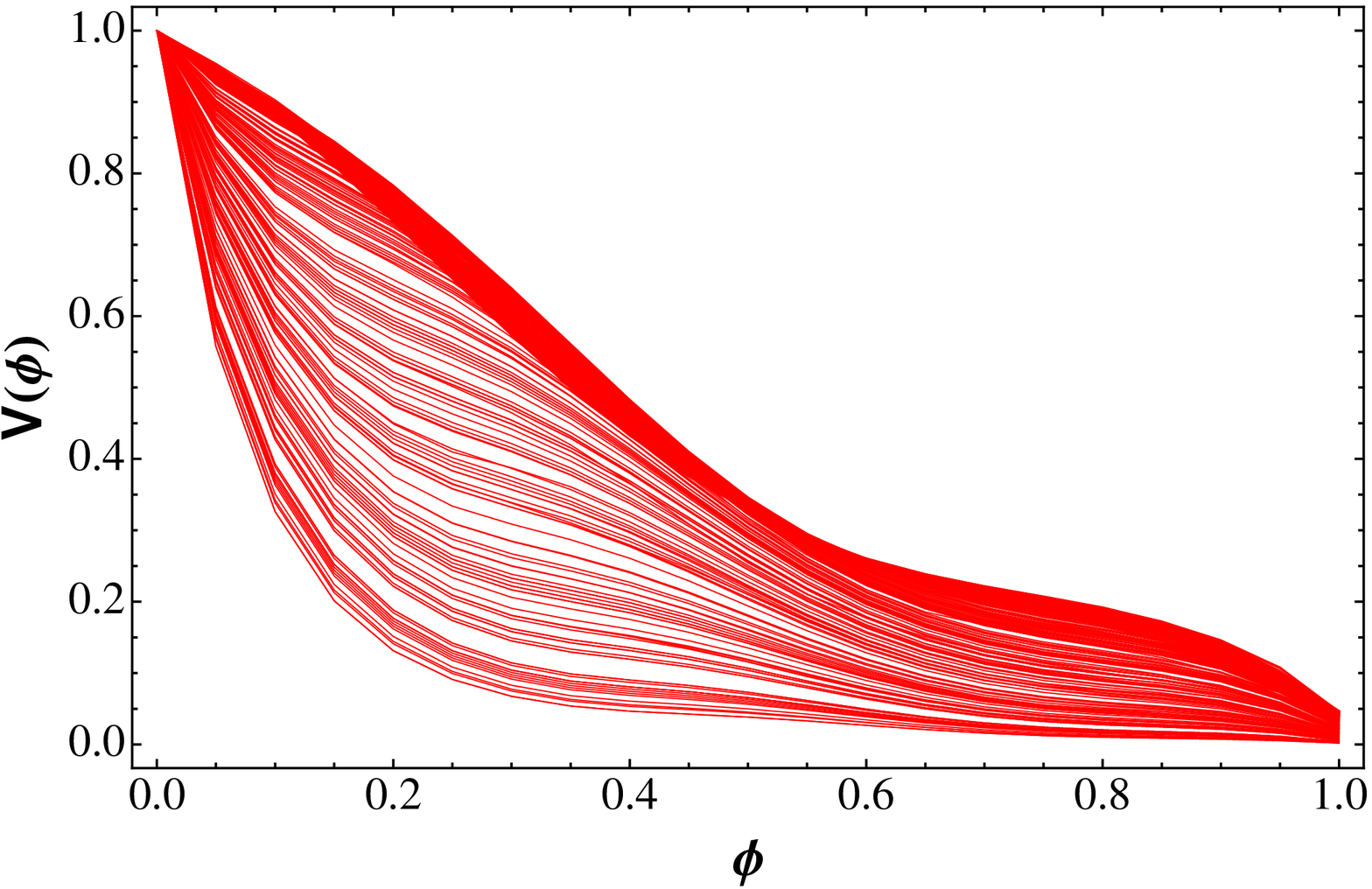}
\includegraphics[width=0.45\linewidth,keepaspectratio,clip]
{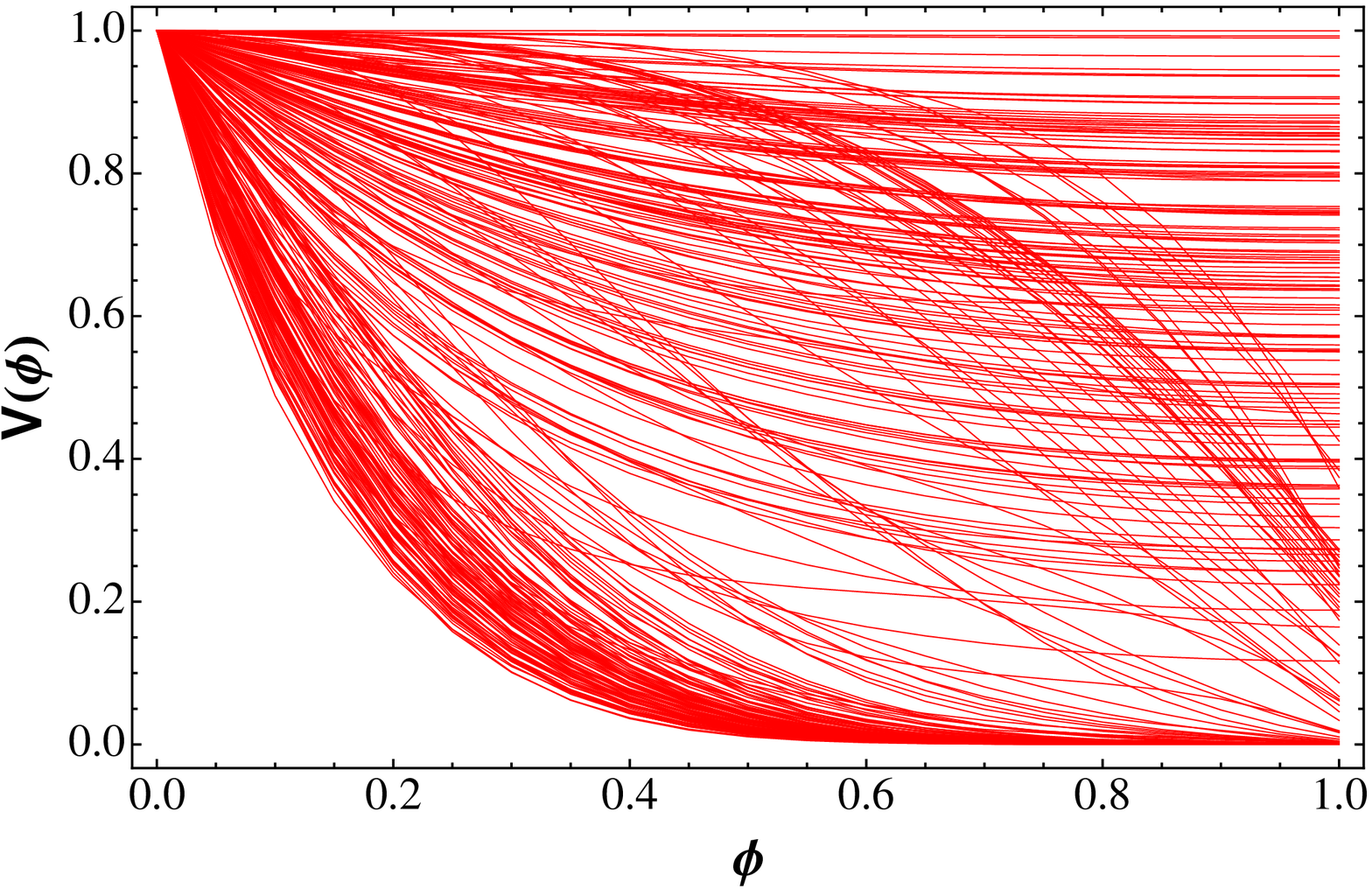}
\includegraphics[width=0.45\linewidth,keepaspectratio,clip]
{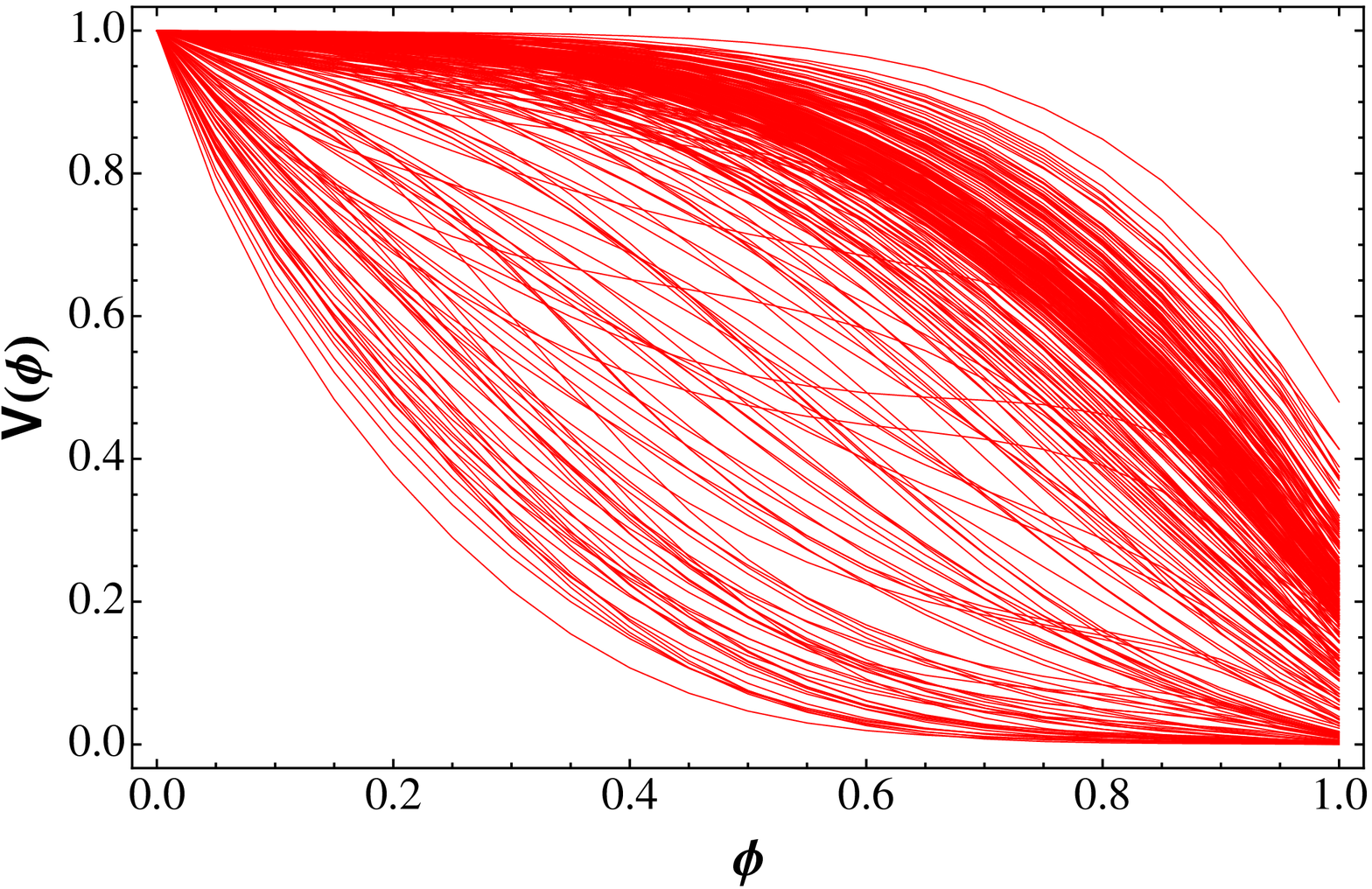}
\caption{Sample of three hundred reconstructed potentials for the four
sets of observational constraints discussed in the text. The
potentials are calculated at the 99\% c.l. All the potentials have been 
rescaled so that $V(\phi = 0) = 1$, and $0\le \phi \le 1$.  Upper left 
panel: WMAP7 with sudden reionization; upper right panel: CMB-ALL with 
sudden reionization; lower left panel: WMAP7 with MH reionization; lower 
right panel: CMB-ALL with MH reionization.}
\label{fig:Vphi1}
\end{center}
\end{figure*}

We have implemented the procedure described above with $N_{MC}=5\times
10^5$. We consider four sets of observational constraints,
corresponding to those obtained in the previous section for the
``WMAP7'' and ``CMB-ALL'' datasets in the two cases of sudden and MH
reionization scenarios. For convenience, we report four sets of
constraints (at the 95\% and 99\% confidence levels) in Table
\ref{table2}. We show a sample of 300 reconstructed potentials in
Fig.\ \ref{fig:Vphi1}.  We have rescaled all the potentials so that
$V(\phi = 0) = 1$ and $0\le \phi \le 1$, so that the figure only
contains information concerning the shape of the inflationary
potential. We see that the WMAP7 data alone do not really constrain the
shape of the inflationary potential, even when more general models of
reionization are considered. However, the situation changes
dramatically when other CMB experiments are included, as it can be
seen from the right panels of Fig.\ \ref{fig:Vphi1}. Presumably this is
due to the fact that when these experiments are included, models with
$n_{run}=0$ are placed outside the observationally allowed region. In
this case, the reionization model makes some difference. In the case
of sudden reionization the possible shapes of the potential are
severely restricted; in particular, it seems that potentials with
$V''(\phi)<0$ are not allowed. When a more general model of
reionization is used, however, more shapes are allowed. This is
probably an result of the fact that the constraints on the scalar
spectral index $n$ are weakened.

Other than the shape of the potential, it is also important to
constrain its amplitude. The reconstruction procedure described above
does not yield the amplitude of the potential; this has to be fixed
from some observational input, like the normalization of the Hubble
parameter. We choose to normalize the Hubble parameter through the
condition on the density contrast:
\begin{equation}
\frac{\delta \rho}{\rho}\simeq \frac{1}{2\pi}\frac{H}{\mpl}
\frac{1}{\sqrt{\epsilon}}\simeq 10^{-5}.
\end{equation}
Once this is done, we find that in all the four cases illustrated in
Table \ref{table2}, $V(\phi)\lesssim 10^{-11}\,\mpl^4$. This
correspond to an upper limit to the energy scale of inflation of
about $10^{16}\,\textrm{GeV}$.

\section{CONCLUSIONS}
\label{sec:concl}

Details of the reionization processes in the late universe are not
very well known. In the absence of a precise, full-redshift evolution
description of the ionization fraction during the reionization period,
a simple parametrization with a single parameter $z_r$ has become
the standard reionization scheme in numerical analyses. More general
reionization schemes have been shown to allow values of the scalar
spectral index consistent with a scale-invariant power spectrum. In
this paper we deduce information about tensor modes, and explore
how the inflation constraints are modified when the standard
reionization assumption is relaxed. The tensor-to-scalar ratio bounds
are largely unmodified under more general reionization
scenarios. Therefore, present (future) primordial gravitational wave
searches are (will be) unaffected by the precise details of
reionization processes. In the absence of a running spectral index,
hybrid models, ruled out in the standard reionization scheme, are
still allowed at the $95\%$ c.l.\ by WMAP7 data.  The constraints on
other inflationary models, such as large-field or small-field models,
do not change. Future Planck data will be able to measure the scalar
spectral index $n$ with unprecedented precision and be sensitive to
tensor modes if $r>0.05$ at the $95\%$ c.l. We also show the impact of
different reionization histories on the reconstruction of the inflaton
potential. The variety of the reconstructed shapes is larger in
extended reionization scenarios than compared to the standard
one. Namely, models with $V''(\phi)<0$ are allowed in general
reionization scenarios. However, the constraints on the amplitude of
the potential remain unchanged: we find an upper bound for the latter
of about $10^{16}\, \textrm{GeV}$, independent of the reionization
process details.

\section*{Acknowledgments}
We would like to thank William Kinney for useful discussion.
O. M. work is supported by the MICINN (Spain) Ram\'on y Cajal
contract, AYA2008-03531 and CSD2007-00060. M. P. is supported by a
MEC-FPU Spanish grant.

\newpage



\end{document}